\documentclass[aps,prd,superscriptaddress,twocolumn,showpacs]{revtex4}
\usepackage{graphicx}
\usepackage{epstopdf}
\usepackage{amsmath}
\usepackage{amsfonts}
\usepackage{amssymb}
\usepackage{latexsym}

\begin{document}
\title{Coulomb's law modification driven by a logarithmic electrodynamics}
\author{Patricio Gaete} \email{patricio.gaete@usm.cl} 
\affiliation{Departamento de F\'{i}sica and Centro Cient\'{i}fico-Tecnol\'ogico de Valpara\'{i}so-CCTVal,
Universidad T\'{e}cnica Federico Santa Mar\'{i}a, Valpara\'{i}so, Chile}
\author{Jos\'{e} A. Helay\"{e}l-Neto}\email{helayel@cbpf.br}
\affiliation{Centro Brasileiro de Pesquisas F\'{i}sicas (CBPF), Rio de Janeiro, RJ, Brasil} 
\author{L. P. R. Ospedal} \email{leoopr@cbpf.br}
\affiliation{Centro Brasileiro de Pesquisas F\'{i}sicas (CBPF), Rio de Janeiro, RJ, Brasil} 
\date{\today}

\begin{abstract}
We examine physical aspects for the electric version of a recently proposed logarithmic electrodynamics, 
for which the electric field of a
point-like charge is finite at the origin. It is shown that this electrodynamics displays the vacuum
birefringence phenomenon in the presence of external magnetic field. Afterwards we compute the
lowest-order modification to the interaction energy by means of the gauge-invariant but path-dependent
variables formalism. 
These are shown to result in a long-range (${1 \mathord{\left/
 {\vphantom {1 {{r^3}}}} \right.\kern-\nulldelimiterspace} {{r^3}}}$-type)
correction, in addition to a linear and another logarithmic correction, to the Coulomb potential. 
\end{abstract}
\maketitle

The subject of quantum vacuum nonlinearities has been of great interest since the work of Heisenberg 
and Euler \cite{H_E}, who showed one of the most astonishing predictions of Quantum Electrodynamics
(QED), namely, the light-by-light scattering in vacuum arising from the interaction of photons with virtual
electron-positron pairs. From then on, the physical consequences of this fundamental
result have been intensely studied from different perspectives \cite{Adler,Costantini,Ruffini,Dunne}. It is to be specially recalled, at this stage, that
only very recently the experimental detection of light-by-light scattering has been reported in the ATLAS
Collaboration \cite{Atlas,Enterria}. It is of interest also to notice experiments related to photon-photon 
interaction physics have suggested that electrodynamics in vacuum is a nonlinear theory 
\cite{Bamber,Burke,nphoton,Tommasini1,Tommasini2}. With this in view, different nonlinear electrodynamics
of the vacuum may have significant contributions to photon-photon scattering such as Born-Infeld  
\cite{BI} and Lee-Wick \cite{Lee-Wick1, Lee-Wick2} theories.

In addition, nonlinear electrodynamics have also attracted considerable attention because they emerge
naturally in string theories. As is well known, the low energy dynamics of D-branes is described by a Born- 
Infeld type action \cite{Tseytlin,Gibbons}. In this perspective, recently nonlinear electrodynamics have been
the object of intensive investigations in the context of phase transitions of black hole physics \cite{Ovgun}.

With these considerations in mind, in previous works \cite{Nonlinear,Logarithmic,Nonlinear2,Nonlinear3}, we
 have
studied the physical effects presented by different models of $(3+1)$-D nonlinear Electrodynamics in
vacuum. In fact, it was shown that for Generalized Born-Infeld, and Logarithmic Electrodynamics the
field energy of a point-like charge is finite. We also point out that Generalized Born-Infeld,
Exponential, Logarithmic and Massive Euler-Heisenberg-like Electrodynamics exhibit the vacuum 
birefringence phenomenon.

Given the ongoing experiments related to light-by-light scattering,
it is of interest to understand better the phenomenological consequences presented by vacuum
electromagnetic nonlinearities. Seem from such a perspective, the present work is an extension of our
previous studies. To do this we consider the electric version of a recently proposed logarithmic
electrodynamics and investigate aspects of birefringence, as well as the computation of the
static potential along the lines of \cite{Nonlinear,Logarithmic,Nonlinear2,Nonlinear3}. In our conventions the signature of
the metric is ($+1,-1,-1,-1$).

Let us start off our considerations with a brief description of the model under consideration (Logarithmic
electrodynamics). In this case, the gauge theory we are considering is described by the Lagrangian density:   
\begin{equation}
{\cal L} =  - 2{\beta ^2}\ln \left[ {1 + \frac{1}{\beta }\sqrt { s {\cal F}} } \right] + 2\beta \sqrt { s {\cal F}}, \label{LNED05}
\end{equation}
where ${\cal F} = \frac{1}{4}F_{\mu \nu } F^{\mu \nu }$. Here, $s=-1$ for ${\left|{\bf E}\right|\geq\left|{\bf B}\right|}$ and $s=1$ for ${\left|{\bf E}\right|{<}\left|{\bf B}\right|}$. Furthermore,
the ${\beta}$ constant has ${\left( {mass} \right)^2}$ dimension in
natural units. We also note that, in a purely electric case, the $\beta$ constant 
could be identified by a background electric field. 

With this, we can write the corresponding equations of motion as
\begin{equation}
\nabla  \cdot {\bf D} = 0, \  \  \
\frac{{\partial {\bf D}}}{{\partial t}} - \nabla  \times {\bf H} = 0, \label{LNED10a}
\end{equation}
\begin{equation}
\nabla  \cdot {\bf B} = 0, \  \  \
\frac{{\partial {\bf B}}}{{\partial t}} + \nabla  \times {\bf E} = 0, \label{LNED10b}
\end{equation}
where the ${\bf D}$ and ${\bf H}$ fields are given by
\begin{equation}
{\bf D}{=}\frac{\sqrt{2}\mathit{\beta}{\bf E}}{\sqrt{2}\mathit{\beta}{+}\sqrt{{-}{s}{(}{\bf E}^{2}{-}{\bf B}^{2}{)}}},\label{LNED15}
\end{equation}
and
\begin{equation}
{\bf H}{=}\frac{\sqrt{2}\mathit{\beta}{\bf B}}{\sqrt{2}\mathit{\beta}{+}\sqrt{{-}{s}{(}{\bf E}^{2}{-}{\bf B}^{2}{)}}}.\label{LNED20}
\end{equation}

Next, we readily see that for an external point-like charge, q, at the origin, the ${\bf D}$-field lies along the radial direction and is
given by ${\bf D} =\frac{q}{{4\pi r^2 }}\hat r$. Hence, for $s=-1$ and $\beta  > 0$, the electrostatic field assumes the form
\begin{equation}
|{\bf E}| = \frac{{\sqrt 2 \beta |q|}}{{4\sqrt 2 \pi \beta {r^2} - |q|}}. \label{LNED25}
\end{equation}
It is  important to observe that this solution is valid for $r > \sqrt {\frac{{|q|}}{{4\sqrt 2 \pi \beta }}}$.

It should be further noted that, for $\beta  < 0$ and $\left( {\sqrt 2 \beta  + |E|} \right) > 0$, the corresponding electrostatic field becomes
\begin{equation}
|{\bf E}| =  - \frac{{\sqrt 2 \beta |q|}}{{4\sqrt 2 \pi \beta {r^2} + |q|}}, \label{LNED25a}
\end{equation}
which is restricted to the domain $\left( {4\sqrt 2 \pi \beta {r^2} + |q|} \right)> 0$, that is,
 for $r > \sqrt {\frac{{|q|}}{{4\sqrt 2 \pi |\beta |}}}$. 

Whereas for $\beta  < 0$ and $\left( {\sqrt 2 \beta  + |E|} \right) < 0$, the electrostatic field reads
\begin{equation}
|{\bf E}| = \frac{{\sqrt 2 \beta |q|}}{{4\sqrt 2 \pi \beta {r^2} - |q|}},  \label{LNED25b}
\end{equation}
which is valid for $r \ge 0$. Evidently, as ${r}\rightarrow{0}$, we get $|{\bf E}| =  - \sqrt 2 \beta$.

In order to explore the optical properties of the model under consideration,
we shall concentrate in the  $s=-1$ and $\beta  > 0$ case.
In a such  case 
\begin{equation}
{\bf D}{=}\frac{\bf E}{{1}{+}\frac{1}{\mathit{\beta}\sqrt{2}}\sqrt{{\bf E}^{2}{-}{\bf B}^{2}}}, \label{LNED30}
\end{equation}
and
\begin{equation}
{\bf H}{=}\frac{\bf B}{{1}{+}\frac{1}{\mathit{\beta}\sqrt{2}}\sqrt{{\bf E}^{2}{-}{\bf B}^{2}}}. \label{LNED35}
\end{equation}

At this point, it is interesting to recall that the complicated field problem can be simplified to a large
 extent if the previous equations are linearized. In this case, we consider a weak electromagnetic wave 
 $({\bf E_p}, {\bf B_p})$ propagating in the presence of a strong constant external field $({\bf E_0}, {\bf 
 B_0})$. Furthermore, for computational simplicity, we will only consider the case of a purely external 
 magnetic field, namely, ${\bf E_0}=0$. We thus find that  
\begin{equation}
{\bf D}{=}\Gamma{\bf E}_{p}, \label{LNED45}
\end{equation}
and
\begin{equation}
{\bf H}{=}\Gamma\left[{{\bf B}_{p}{-}\frac{\left({{\bf B}_{p}\cdot{\bf B}_{0}}\right)}{{\mathit{\beta}}^{2}\left({{1}{-}\frac{{\bf B}_{0}^{2}}{2{\mathit{\beta}}^{2}}}\right)}{\bf B}_{0}}\right], \label{LNED50} 
\end{equation}
with
\begin{equation}
\Gamma  = \left( {1 - \frac{{{\bf B}_0^2}}{{2{\beta ^2}}}} \right)\left( {1 - \frac{i}{{\sqrt 2 \beta }}\sqrt {{\bf B}_0^2} } \right),
\label{LNED55} 
\end{equation} 
where we have keep only linear terms in ${\bf E_p}$, ${\bf B_p}$. From these expressions we readily 
deduce that 
\begin{equation}
{\varepsilon _{ij}} = \Gamma {\delta _{ij}} , \label{LNED60a}
\end{equation}
and
\begin{equation}
{\left({{\mathit{\mu}}^{{-}{1}}}\right)}_{ij}{=}\Gamma\left({{\mathit{\delta}}_{ij}{-}\frac{1}{{\mathit{\beta}}
^{2}\left({{1}{-}\frac{{\bf B}_{0}^{2}}{2{\mathit{\beta}}^{2}}}\right)}{B}_{0i}{B}_{oj}}\right). \label{LNED65b}
\end{equation}

Next, we make a plane wave decomposition for the fields ${\bf E_p}$ and ${\bf B_p}$:
\begin{equation}
{{\bf E_p}}\left( {{\bf x}
,t} \right) = {\bf E}
{e^{ - i\left( {wt - {\bf k} \cdot {\bf x}} \right)}}, \ \ \
{{\bf B_p}}\left( {{\bf x},t} \right) = {\bf B}{e^{ - i\left( {wt - {\bf k} \cdot {\bf x}} \right)}}. \label{LNED70}
\end{equation}
Assuming then that the external magnetic field is in the direction $z$, ${\bf B_0}  = B_0 {\bf e}_3$, and
the light wave moves along the $x$ axis, the corresponding Maxwell equations assume the form
\begin{equation}
\left( {\frac{{{k^2}}}{{{w^2}}} - {\varepsilon _{22}}{\mu _{33}}} \right){E_2} = 0, \label{LNED75a}
\end{equation}  
and
\begin{equation}
\left( {\frac{{{k^2}}}{{{w^2}}} - {\varepsilon _{33}}{\mu _{22}}} \right){E_3} = 0. \label{LNED75b}
\end{equation} 
In passing we note that the preceding equations were obtained in the limit ${{\bf B}_0} \gg {{\bf B}_p}$ and $\beta  \gg |{{\bf B}_0}|$.

Accordingly, two interesting situations arise from the foregoing equations.

First, if ${\bf E}\ \bot \ {\bf B}_0$ (perpendicular polarization), from (\ref{LNED75b}) $E_3=0$, and from
 (\ref{LNED75a}) we get $\frac{{{k^2}}}{{{w^2}}} = {\varepsilon _{22}}{\mu _{33}}$. The
 dispersion relation therefore has the form 
\begin{equation}
{n_ \bot } = \sqrt {\frac{{1 - \frac{{{\bf B}_0^2}}{{2{\beta ^2}}}}}{{1 - \frac{{3{\bf B}_0^2}}{{2{\beta ^2}}}}}}. \label{LNED80}
\end{equation}
Second, if ${\bf E}\ || \ {\bf B}_0$ (parallel polarization), from (\ref{LNED75a}) $E_2=0$, and from
 (\ref{LNED75b}) we get $\frac{{{k^2}}}{{{w^2}}} = {\varepsilon _{33}}{\mu _{22}}$. The dispersion 
 relation reduces to
\begin{equation}
{n_\parallel } = 1.  \label{LNED85}
\end{equation}
Thus we have the vacuum birefringence phenomenon, that is, electromagnetic waves with different
 polarizations have different velocities. It is of interest also to notice that for the $s=-1$ and $\beta  < 0$ case, 
 we obtain the same dispersion equations as expressed by eqs. (\ref{LNED80}) and (\ref{LNED85}).

We shall now compute the interaction energy between static point-like sources for this logarithmic
 electrodynamics. The starting point is the Lagrangian density ($s=-1$, $\beta  > 0$):
\begin{equation}
{\cal L} = \left( { - {\cal F}} \right) - \frac{2}{{3\beta }}{\left( { - {\cal F}} \right)^{{3 \mathord{\left/
 {\vphantom {3 2}} \right.\kern-\nulldelimiterspace} 2}}}. \label{LNED90}
\end{equation}
To get the last line we used $\beta  \gg \sqrt { - {\cal F}}$, according to our preceding development. We remark that the new feature of the present model is the non-trivial presence of the exponent ${3 \mathord{\left/
 {\vphantom {3 2}} \right.\kern-\nulldelimiterspace} 2}$ in expression (\ref{LNED90}). Thus, the purpose of analyzing this model here is to investigate the impact of this exponent on a physical observable.

As was explained in \cite{Nonlinear,Logarithmic,Nonlinear2,Nonlinear3}, an alternative way of writing the above equation is making use of an auxiliary field $v$, such that its equation of motion gives back the original equation. We thus find that equation (\ref{LNED90}) can be brought to the form
\begin{equation}
 {\cal L}{=}\left({{1}{-}\frac{2}{\mathit{\beta}}{v}}\right)\left({{-}{\cal F}}\right){+}\frac{8}{{3}\mathit{\beta}}{v}^{3}. \label{LNED95}
\end{equation}
It is also convenient to rewrite this equation as
\begin{equation}
{\cal L} =  - \frac{1}{4}\frac{1}{V}{F_{\mu \nu }}{F^{\mu \nu }} -  
\frac{{\mathit{\beta}}^{2}}{3}\frac{{\left({{V}{-}{1}}\right)}^{3}}{{V}^{3}}, \label{LNED100}
\end{equation}
where we have used $\frac{1}{V}{=}\left({{1}{-}\frac{2}{\mathit{\beta}}{v}}\right)$.
Here, the quantization is carried out using Dirac's procedure. The canonically conjugate momenta are $    
{\Pi ^\mu } =  - \frac{1}{V}{F^{0\mu }}$. In this manner we have two primary constraints 
$\Pi ^0  = 0$ and $p \equiv \frac{{\partial L}}{{\partial \dot V}} = 0$. Furthermore, ${\Pi _i} = \frac{1}{V}{E_i}$. The canonical Hamiltonian is then
\begin{eqnarray}
{H_C} &=& \int {{d^3}x} \left\{ {{\Pi _i}{\partial ^i}{A_0} + \frac{V}{2}{{\bf \Pi} ^2} + \frac{1}{{2V}}{{\bf B}^2}}
 \right\} \nonumber\\
 &-&\frac{{\mathit{\beta}}^{2}}{3}\int{{d}^{3}x}\frac{{\left({{V}{-}{1}}\right)}^{3}}{{V}^{3}}. \label{LNED105}
\end{eqnarray}

Temporal conservation of the primary constraint, $\Pi^{0}$, leads to the secondary constraint $\Gamma _1  = \partial _i \Pi ^i  = 0$. Whereas for the constraint $p$, we obtain the auxiliary field $V$  
\begin{equation}
V={-}\frac{\sqrt{2}\mathit{\beta}}{2\sqrt{{{\bf \Pi}}^{2}}}\left({{1}{+}\sqrt{{1}{+}\frac{4}{\sqrt{2}\mathit{\beta}}
\sqrt{{{\bf \Pi}}^{2}}}}\right), \label{LNED110}
\end{equation}
which will be used to eliminate $V$. It is worthwhile mentioning that to get this last expression we have
 ignored the magnetic field in equation (\ref{LNED105}), because it add nothing to the static potential 
 calculation.

By proceeding in the same way as in \cite{Nonlinear,Logarithmic,Nonlinear2,Nonlinear3}, we obtain the extended Hamiltonian as
\begin{eqnarray}
{H} &=& \int {{d^3}x} \left\{ {w(x){\partial ^i}{\Pi _i} + \frac{V}{2}{{\bf \Pi} ^2} } \right\} \nonumber\\
&-&\frac{{\mathit{\beta}}^{2}}{3}\int{{d}^{3}x}\frac{{\left({{V}{-}{1}}\right)}^{3}}{{V}^{3}}, \label{LNED115}
\end{eqnarray}
where $w(x)$ is an arbitrary Lagrange multiplier and $V$ is given by (\ref{LNED110}).

We can now compute the interaction energy for the model under consideration. To accomplish this task,
 we shall recall that the interparticle potential energy can be calculated through the  expression 
 \cite{Nonlinear,Logarithmic,Nonlinear2,Nonlinear3}
\begin{equation}
{\cal V} \equiv e\left( {{\cal A}_0 \left( {\bf 0} \right) - {\cal A}_0 \left( {\bf L} \right)} \right), \label{LNED125}
\end{equation}
where the physical scalar potential is given by
\begin{equation}
{\cal A}_0 (t,{\bf r}) = \int_0^1 {d\lambda } r^i E_i (t,\lambda
{\bf r}). \label{LNED130}
\end{equation}
This equation follows from the vector gauge-invariant field expression
\begin{equation}
{\cal A}_\mu  (x) \equiv A_\mu  \left( x \right) + \partial _\mu  \left( { - \int_\xi ^x {dz^\mu  A_\mu  \left( z \right)} } \right), \label{LNED135}
\end{equation}
where the line integral is along a spacelike path from the point $\xi$ to $x$, on a fixed slice time. It should again be stressed here that the gauge-invariant variables (\ref{LNED135}) commute with the sole first constraint (Gauss law). We have skipped all the technical details and refer to \cite{Nonlinear,Logarithmic,Nonlinear2,Nonlinear3} for them. 

In the presence of an external current $J^0$, we first observe that Gauss' law (obtained from the 
Hamiltonian formulation above) reduces to
\begin{equation}
{\partial _i}{\Pi ^i} = {J^0},  \label{LNED140}
\end{equation}
where ${E^i} =V{\Pi ^i} $ and $V$ is given by equation (\ref{LNED110}). For ${J^0}({\bf r}) = e{\delta 
^{\left( 3 \right)}}\left( {\bf r} \right)$, the electric field is then   
\begin{equation}
{\bf E}{=}{-}\frac{\sqrt{2}\mathit{\beta}}{2}{\left({{1}{+}\sqrt{{1}{+}\frac{e}{\mathit{\pi}\sqrt{2}\mathit{\beta}}\frac{1}{{r}^{2}}}}\right)}\hat{r}. \label{LNED145}
\end{equation}

With the aid of equation (\ref{LNED145}), equation (\ref{LNED130}) can be written as
\begin{equation}
{\cal A}_{0}\left({t,{\bf r}}\right){=}\frac{\sqrt{2}\mathit{\beta}}{2}\mathop{\int}\nolimits_{0}\nolimits^{r}{dz}
\left({{1}{+}\sqrt{{1}{+}\frac{e}{\mathit{\pi}\sqrt{2}\mathit{\beta}}\frac{1}{{z}^{2}}}}\right), \label{LNED150}
\end{equation}
From this last equation it follows that 
\begin{eqnarray}
{\cal A}_{0}\left({t,{\bf r}}\right)&=&\frac{\sqrt{2}\mathit{\beta}}{2}{r} \nonumber\\
&+&\frac{\sqrt{2}\mathit{\beta}}{2}\left\{{\sqrt{{p}{+}{r}^{2}}{+}\sqrt{p}{ln}\left[{\frac{r}{{p}{+}\sqrt{p}\sqrt{{p}{+}{r}^{2}}}}\right]}\right\}, \nonumber\\
\label{LNED155}
\end{eqnarray}
with ${p}{=}\frac{e}{\mathit{\pi}\sqrt{2}\mathit{\beta}}$.

Making use of the foregoing equation, we finally obtain the potential (to the lowest order  in $\beta$) for
 a pair of static point-like opposite charges located at $\bf 0$ and $\bf L$, 
\begin{eqnarray}
{\cal V}&=&{-}\frac{{e}^{2}}{{4}\mathit{\pi}}\frac{1}{r}{+}\frac{{e}^{3}\sqrt{2}}{32{\mathit{\beta}\mathit{\pi}}^{2}}\frac{1}{{r}^{3}}{-}{e}\sqrt{2}\mathit{\beta}{r} \nonumber\\
&-&\sqrt{\frac{\mathit{\beta}}{\mathit{\pi}}}\frac{{e}^{\mbox{\tiny $\raise0.7ex\hbox{${5}$}\!\!\left/{}\right.\!\!\lower0.7ex\hbox{${2}$}$}}}{{2}^{\mbox{\tiny $\raise0.7ex\hbox{${5}$}\!\!\left/{}\right.\!\!\lower0.7ex\hbox{${4}$}$}}}{ln}\left[{\frac{r}{\sqrt{\frac{e}{\sqrt{2}\mathit{\pi}\mathit{\beta}}}\left({{r}{+}\sqrt{\frac{e}{\sqrt{2}\mathit{\pi}\mathit{\beta}}}}\right)}}\right],  \label{LNED160}
\end{eqnarray}
after subtracting a self-energy term, and   ${r}{=}\left|{\bf L}\right|$. Here, an interesting matter comes out. Although the third term
in the previous equation is proportional to $r$, such a term has the wrong sign. Therefore we cannot speak about confinement. 
This is a consequence of the non-trivial presence of the exponent ${3 \mathord{\left/
 {\vphantom {3 2}} \right.\kern-\nulldelimiterspace} 2}$ in expression (\ref{LNED90}). It is also important to observe that for the $s=-1$ and $\beta  < 0$ case, we obtain the same static potential profile as expressed by eq. (\ref{LNED160}).\\

In order to illustrate this last point, we shall consider the model defined by the following Lagrangian
 density
\begin{equation}
{\cal L}{=}{-}{2}{\mathit{\beta}}^{2}\ln\left[{{1}{+}\frac{1}{\mathit{\beta}}\sqrt{{-}{\cal F}}}\right]. \label{FR05}
\end{equation}

Following our earlier procedure, we first observe that for a point-like charge, e, at the origin, the 
electrostatic field is given by
\begin{equation}
\left|{\bf E}\right|{=}\sqrt{2}\mathit{\beta}\left({{1}{-}\frac{{4}\mathit{\pi}}{\sqrt{2}e}{r}^{2}}\right), 
\label{FR10}
\end{equation}
hence, as ${r}\rightarrow{0}$, we get $\left|{\bf E}\right|{=}\sqrt{2}\mathit{\beta}$.

Again, to leading order in $\beta$, equation (\ref{FR05}) becomes
\begin{equation}
{\cal L}{=}\left({{-}{\cal F}}\right){-}{2}\mathit{\beta}\sqrt{{-}{\cal F}}. \label{FR15}
\end{equation}
By proceeding in the same way as before, we obtain the static potential for two opposite charges located at $\bf 0$ and $\bf L$ turns out to be 
\begin{equation}
{\cal V}{=}{-}\frac{{e}^{2}}{{4}\mathit{\pi}{r}}{+}\frac{\mathit{\beta}{e}}{{2}\mathit{\pi}}{r}. \label{FR25}
\end{equation}
Expression (\ref{FR25}) immediately shows the effect of the exponent being now ${1 \mathord{\left/
 {\vphantom {1 2}} \right.\kern-\nulldelimiterspace} 2}$ (expression (\ref{FR15})) on the static potential, which is the sum of a Coulomb and a linear potential, leading to the confinement of static charges.
Interestingly enough, the above static potential profile is analogous to that encountered in a new nonlinear electrodynamics governed by the Lagrangian density
\begin{equation}
{\cal L}=\beta^{2}\left\{{1-{\left[{1+{{2\sqrt{2}}\over{\beta}}\sqrt{-{\cal F}}}\right]}^{p}}\right\}, \label{FR30}
\end{equation}
with $0 < p < 1$. For $p= {1 \mathord{\left/{\vphantom {1 2}} \right.\kern-\nulldelimiterspace} 2}$ we obtain the same potential as the one given by equation (\ref{FR25}).

We conclude by putting our work in its proper perspective. This paper is a sequel to \cite{Nonlinear,Logarithmic,Nonlinear2,Nonlinear3}, where once again we
have exploited a correct identification of field degrees of freedom with observable quantities. It was shown that in
this new electrodynamics the phenomenon of birefringence takes place in the presence of external magnetic fields. 
Subsequently we have studied the interaction energy. Our analysis reveals that the static potential profile contains
in a long-range (${1 \mathord{\left/{\vphantom {1 {{r^3}}}} \right.\kern-\nulldelimiterspace} {{r^3}}}$-type)
correction, in addition to a linear and another logarithmic correction, to the Coulomb potential.

One of us (P. G.) was partially supported by Fondecyt (Chile) grant 1180178 and by Proyecto Basal FB0821.  LPRO is grateful
to the National Council for Scientific and Technological Development (CNPq/MCTIC) for financial support through the PCI-DB funds.

\end{document}